\documentclass{elsart3}
\usepackage{times}

\usepackage{graphicx}
\usepackage{psfig}

\usepackage{amssymb}
\begin{document}

\begin{frontmatter}


\title{Coupled Electron Ion Monte Carlo Calculations of Atomic Hydrogen }

% Attention please ! do not leave blank lines here !
\author[ia1]{Markus Holzmann\corauthref{cor1}}
\corauth[cor1]{Corresponding Author:}
\ead{markus@lptl.jussieu.fr}
\author[ia2]{Carlo Pierleoni}
\author[ia3]{David M. Ceperley}
\address[ia1]{LPTL, UMR 7600 of CNRS, Universit\'e P. et  M. Curie, Paris, France}
\address[ia2]{INFM and Department of Physics, University of L'Aquila, Via Vetoio, I-67010 L'Aquila, Italy}
\address[ia3]{Physics Department, University of Illinois at Urbana-Champaign, Urbana, IL 61801, USA}
%\thanks[label1]{On leave of absence from XXX }

\begin{abstract}
% Text of abstract
We present a new Monte Carlo method which couples Path Integral
for finite temperature protons with Quantum Monte Carlo for ground
state electrons, and we apply it to metallic hydrogen for
pressures beyond molecular dissociation. This method fills the gap
between high temperature electron-proton Path Integral and ground
state Diffusion Monte Carlo methods. Our data exhibit more
structure and higher melting temperatures of the proton crystal
than Car-Parrinello Molecular Dynamics results using LDA. We
further discuss the quantum motion of the protons and the zero
temperature limit.

%Put here the text of your abstract. Max 3000 characters.
\end{abstract}

\begin{keyword}
Quantum Monte Carlo\sep metallic hydrogen 
\PACS 05.30.Lq \sep 71.10.+x \sep 64.30.+t \sep 02.70.Lq 
\end{keyword}
\end{frontmatter}

% main text
\section{Introduction}
\label{sec1} Quantum Monte Carlo (QMC) methods have been developed
for accurately solving the many-body Schr\"odinger equation. Zero
temperature Variational Monte Carlo (VMC)  and Diffusion Monte
Carlo (DMC), and finite temperature  Path Integral Monte Carlo
(PIMC) are currently the most accurate and general methods for
computing static properties of a quantum system
\cite{foulkes01,rmp95}. They have been successfully applied to
simple quantum many-body systems, such as the electron gas,
hydrogen, and helium.

Recently, there have been new attempts\cite{dewing01,dmc03,pch04}
to calculate properties of disordered systems such as liquid
hydrogen within QMC. For this system, VMC/DMC  and PIMC are
computationally too inefficient to provide definite answers, e.g.
regarding the nature of the melting transition from liquid to
solid or metal to insulator. Whereas PIMC calculations have been
done at comparatively high temperatures \cite{pimc}, this method
becomes computationally inefficient at temperatures lower than
roughly $1/20$ of the Fermi temperature. Zero temperature
calculations (VMC, DMC) have been used for ground state
calculations where both electronic and protonic degrees of freedom
are treated quantum mechanically \cite{dmc81,natoli93}. However, the
convergence of these calculations suffers from the different
masses of protons and electrons which introduce two time scales
differing by three orders of magnitude, and, more important, low
temperature properties are inaccessible by these ground state
methods. To fill this gap, the Coupled Electron-Ion Monte Carlo
(CEIMC) has been developed \cite{dewing01,dmc03} to combine a
classical or quantum Monte Carlo simulation of the nuclei at
non-zero temperature with a QMC calculation for the electronic
energies where the Born-Oppenheimer approximation helps to
overcome the time scale problem.

In Ref.~\cite{pch04}, the CEIMC method has been applied to
determine the equation of state of hydrogen for temperatures
across the melting of the proton crystal. More structure and
higher melting temperatures of the proton crystal compared to
Car-Parrinello Molecular Dynamics (CPMD) results using LDA\cite{kh95} have
been found. In this paper, we shortly summarize the method and the
results as reported in Ref.~\cite{pch04} and discuss in more
detail the quantum effects of the protons \cite{marx,natoli-th}.

\section{Method}

In the CEIMC method, the proton degrees of freedom are advanced by
a Metropolis algorithm in which the energy difference between the
actual state $S$ and the trial state $S'$ is computed by a Quantum
Monte Carlo calculation 
%(either variational (VMC) or by Reptation
%Monte Carlo (RQMC) \cite{BaroniMoroni}). 
The energies of the
states are calculated within the Born-Oppenheimer approximation,
where the electrons are assumed to remain in the ground state with
respect to the actual protonic positions. Since the
Born-Oppenheimer energies $E(S)$  and $E(S')$ have to be sampled
by a QMC calculation, they are affected by statistical noise which
would bias the Monte Carlo sampling of the protons. At first sight
one might expect that for an unbiased calculation one will need to
reduce the accuracy of the energy difference $E(S)-E(S')$ much
below $k_BT$. However, it has been shown that unbiased sampling of
the proton configurations can be efficiently achieved by using the
penalty method\cite{penalty}, a generalization of the Metropolis
algorithm, where detailed balance is satisfied on average.

Since only differences of electronic energies are needed, we
sample the electronic degrees of freedom according to the sum of
the electronic distribution functions ({\it e. g.} the square of
the trial wave function in VMC) for the $S$ and $S'$ states, and
we compute the energies for the two states as correlated sampling
averages\cite{dewing01,dmc03}. For the typical size of the proton
moves (between 0.01\AA and 0.5\AA for classical protons)
%depending on  density and temperature)
%and the typical system size
this method
is much more efficient than performing two independent
electronic calculations \cite{dewing01,dmc03}.

An essential part of the CEIMC method is the choice of the trial
wavefunction needed to calculate the Born-Oppenheimer energies.
Variational Monte Carlo depends crucially on the quality of the
trial wavefunction. To go beyond VMC, we implemented a Reptation
Quantum Monte Carlo algorithm (RQMC)\cite{BaroniMoroni} to sample
more accurately the electronic ground state. Similar to DMC, RQMC
projects the trial wavefunction on to the ground state within the
Fixed-Node approximation.
%The imaginary time propagator is the usual importance sampling
%Green's function of the DMC\cite{rmp95} times the exponential of
%the time integral of the local energy over the path, and projects
%the trial function onto the ground state for large imaginary time.
%In this limit the center of the paths are distributed according to
%the square of the ground state wave function appropriate to the
%fixed node constraint.
A high quality trial wave functions is important to relax to the
ground state with a very limited number of time slices and to
provide accurate nodes. RQMC, being a Metropolis based method, is
more easily used to compute energy differences; conversely, the
correlated sampling method within DMC is more involved because of
the branching step.

To reduce finite size effects in
metallic systems, we average over
twisted boundary conditions (TABC) when computing electronic energies
within CEIMC
({\it i.e.} we integrate over the Brillouin zone
of the super cell)\cite{lzc01,dmc03}.
%and averaged the electronic quantities
%Since we are dealing with metallic systems, we have implemented
%twist averaged boundary conditions
%({\it i. e.} integration over the Brillouin zone
%of the super cell) %and averaged the electronic quantities
%over the twist phase (a three dimensional phase) in order
%to reduce finite size effects (TABC) \cite{lzc01,dmc03}.
%All properties are averaged
%over 1000 different k-points.
%(actually 500 values because of the inversion symmetry:
%${\bf \theta}\rightarrow -{\bf\theta}$).
%For each pair of proton configurations and each phase
%value we perform an independent calculation and we take advantage
%of the phase average to reduce the noise on the energy
%difference. For a proposed protonic move, the electronic
%configuration of each phase needs to be relaxed before computing
%the energy difference.
%For the typical protonic displacement, we
%have found that the typical correlation time of the energy
%difference is 10 electronic steps in VMC so that we discard the
%first 30 electronic steps and
%compute the energy difference over
%100 electronic steps/k-point.
%After averaging over k-points, the noise level is
%small enough to simulate temperatures as low as 100K\cite{dmc03}.

Quantum effects for protons are relevant at high pressure.
We represent protons by imaginary time path integrals without
considering the statistics of the protons.
%Protons are at finite temperature and the proton imaginary time runs over the
%inverse temperature axis.
(those effects are negligible in this temperature-density range.)
%the effects of quantum statistics for the protons are negligible
%so we use Boltzmann statistics.
%Since we have computed
%only averages of diagonal operators in the position
%representation, each quantum proton corresponds to a closed path.
For efficiency, it is important
%essential to choose a suitable Trotter break-up of the protonic
%action
to minimize the number of protonic time slices. We have used the
pair action of an effective proton-proton potential and treated
the difference between the true Born-Oppenheimer energy and the
effective potential with the primitive approximation\cite{rmp95}.
%With this action, we find that a proton imaginary time step
%$\tau_p=0.3\times10^{-3} K^{-1}$ is appropriate for $r_s\ge 1$ so
%that few tens of time slices allow for calculations above 100K.
When coupled with TABC, rather than using all the k-points for
each protonic time slice, we can, randomly assign a subset of
k-points to each protonic slice without introducing a detectable
systematic effect.

\section{Results}
\subsection*{Comparison of CEIMC with CPMD}
We first consider classical protons. For classical protons it is
possible to compare the CEIMC results with previous Car-Parrinello
molecular dynamics (CPMD) \cite{kh95}, the only difference is the
method to calculate the potential energy surface. Whereas in CEIMC
the Born-Oppenheimer energies are calculated by QMC methods, CPMD
uses density functional theory (DFT)
%within the local density approximation (LDA)
to calculate electronic energies. Both methods
are in principle exact but rely on approximations of the unknown
nodes of the trial wavefunctions in QMC and on the approximation
of the unknown exchange-correlation energy functional in DFT. In
Fig.~1 we compare the proton correlation function $g_{pp}(r)$ of
both methods. The CEIMC results show more structure than CPMD of
Ref.~\cite{kh95} using LDA. A better agreement is observed when
CPMD results at temperature $T$ are compared to CEIMC results at
temperature $2T$ for $300 \le T \le 3000$. Already previous
studies suggested that the Born-Oppenheimer energy surface from
DFT-LDA is flatter than the more accurate one from QMC. In
Ref.~\cite{natoli93} differences in energy among various crystal
structures obtained within LDA were found smaller than DMC
energies by roughly a factor of two. Further, in
Ref.~\cite{dewing01,dmc03}, the energies for random configurations
of molecular hydrogen showed more variation within DMC than LDA.
\begin{figure}
\centerline{\psfig{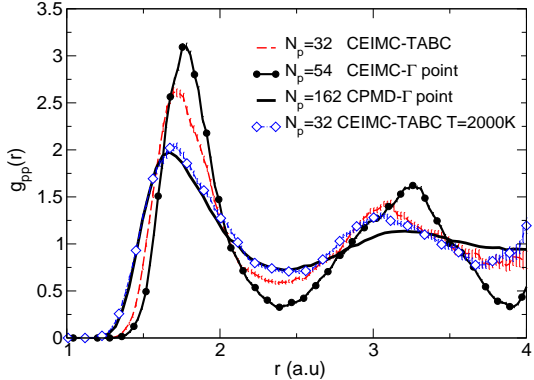}}
\caption{Pair correlation function at $r_s=1,
T=1000K$. Comparison between CEIMC-VMC-TABC with 32 protons,
CEIMC-VMC-PBC with 54 protons and CPMD-LDA with 162 protons
(simulation with $N_p=54$ provides identical correlation). Data
from CEIMC-VMC-TABC at T=2000K (stars) are also reported. }
\label{fig:lda}
\end{figure}

Note that the CEIMC curves of Fig.~1 are all obtained using VMC to
obtain the electronic Born-Oppenheimer energies. RQMC requires
roughly one order of magnitude more computer time then VMC. For
this reasons large part of our results are based on VMC
calculation and RQMC is only exploited to estimate the systematic
error of VMC. For the system with $54$ protons at the $\Gamma$
point, we have found no detectable differences in the correlation
function between VMC and RQMC.

\subsection*{The Quantum effects of the protons}
The quantum effects of the protons are summarized in
Fig.~2 which shows the kinetic energy of the protons versus temperature
for three different densities ($r_s=1.2, 1.0, 0.8$) and the deviation from
its classical value $3 k_B T/2$.
\begin{figure}
\centerline{\psfig{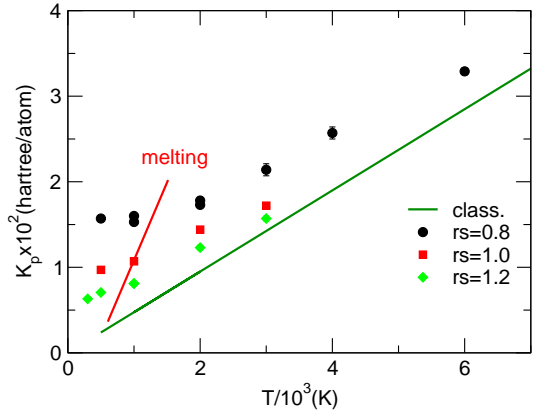}}
\caption{CEIMC-VMC-TABC with 54 protons. Protonic
kinetic energy per particle at various densities versus
temperature. The red line estimates the melting of the bcc crystal
from the Lindemann ratios. } \label{fig:kp}
\end{figure}
Furthermore, Fig.~3 compares $g_{pp}(r)$ for a classical and  a
quantum calculation which directly shows that the zero point
motion is extremely important. The quantum effects not only affect
the proton kinetic energy but also increase the electronic
energies and the configuration energy, at least at the variational
level used in the present calculations.
\begin{figure}
\centerline{\psfig{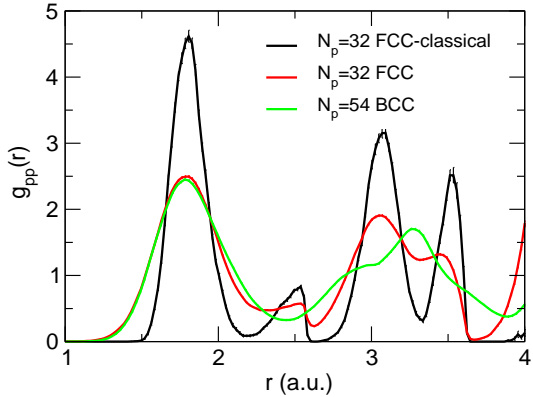}}
\caption{Comparision of $g_{pp}(r)$ between classical and quantum
protons at $r_s=1$ and $T=500K$ (CEIMC-VMC-TABC). } \label{fig:grpp}
\end{figure}

\subsection*{Trial wave functions and zero temperature results}
In this subsection we discuss the zero temperature limit by
VMC/DMC calculations at $T=0$ of dynamical protons. In most of our
CEIMC calculations, analytic trial wave functions including
backflow and three-body correlation\cite{hcpe03} for the
electronic degrees of freedom have been used. Here, we use this
important ingredient of CEIMC to test their accuracy by comparing
their energies for ``dynamic'' protons with the results of Natoli
et al.\cite{natoli-th} in the ground state using LDA trial
wavefunctions. In these calculations, the LDA trial function was
computed for a perfect bcc lattice and then modified for use within
DMC calculations of moving protons in order to avoid recalculation
of the LDA orbitals. In the present calculations, in addition to
the backflow-threebody wavefunction for the electronic part of the
wavefunction, the protonic part of the wavefunction contains a
Jastrow correlation and has a Gaussian orbital localizing it to
the bcc lattice sites (that was also in the Natoli calculation);
exchange effects for protons due to statistics are neglected. In
contrast to CEIMC,  these DMC calculations do not make use of the
Born-Oppenheimer approximation. From table~1 we see that the
backflow wavefunctions have a lower energy by 2-3mH/atom within
VMC and 2mH/atom within DMC. However, the analytical functions are
particularly appropriate to our methods since their computational
cost is much less then solving the Kohn-Sham equations for a
random arrangement of 50 to 100 protons.

%\begin{table}
%\begin{tabular}{|c|c|l|l|l|} \hline
% N   & wavefunction & $E_v$& $\sigma^2$ &$E_{DMC}$ \\\hline
% 16    &   LDA      &   -0.4870 (10)&           & -0.4890 (5)   \\
%     &   BF-A & -0.4878 (1)& 0.0181 (4) & -0.4905(1) \\\hline
%  54   &   LDA    &    -0.5365 (5) &             & -0.5390 (5)   \\
%     &   BF-A & -0.5353 (2) & 0.0178 (2) & -0.5382(1) \\\hline
%128    &   LDA    &   -0.4962  (2) &             & -0.4978 (2)   \\
%     &   BF-A &  -0.4947(2) & 0.023(1) & -0.4978(4) \\\hline
%\end{tabular}
%\caption{ Energies for bcc hydrogen at $r_s$=1.31.
%LDA means LDA orbitals times an optimized 1 body factor
%and Jastrow factor\protect\cite{natoli-th}, BF-A are the analytical
%wavefunctions using backflow.
% Energies are given in
%hartrees per atom. Periodic boundary conditions  ($\Gamma$ point)
%and Ewald sums were used. $\sigma$ is the variance per electron.
%}
%\end{table}
\begin{table}
\begin{tabular}{|c|c|l|l|l|} \hline
 N   & wavefunction & $E_v$& $\sigma^2$ &$E_{DMC}$ \\\hline
 16    &   LDA      &   -0.4678 (2)&           &    \\
     &   BF-A & -0.4724(1) & 0.030 (2) &  \\\hline
  54   &   LDA    &    -0.5195 (3) &             & -0.52415 (5)   \\
     &   BF-A & -0.52194 (5) & 0.025(1)  & -0.52610(7) \\\hline
\end{tabular}
\caption{ VMC and DMC Energies in h/atom (and variance per electron
$\sigma$) for dynamic protons and electrons
at $r_s$=1.31 at $T=0$. LDA means LDA orbitals times an optimized
1 body factor and Jastrow factor\protect\cite{natoli93} used as
trial wavefunction, BF-A are the analytical wavefunctions using
backflow times an optimized gaussian for the protons. 
}
\end{table}

To extend the data of the equation of state of Ref.~\cite{pch04}
to the zero temperature limit, we performed VMC-TABC calculations
at zero temperature. The results are summarized in table 2. The
zero point energy is obtained by subtracting the energy in a
static bcc lattice of protons from the energy of the dynamic
electron-proton system. In the harmonic approximation, half of the
zero point energy will contribute to the potential and the other
half to the kinetic energy of the protons at zero temperature and
we can compare with the protonic kinetic energies calculated in
Ref. \cite{pch04}. The zero temperature kinetic energies of the
protons obtained is systematically higher than the zero
temperature extrapolated results of Ref.\cite{pch04}. It is
known\cite{dmc81,natoli93} that anharmonic effects are large for
high pressure hydrogen.  The difference could also be due to
limitations of the gaussian trial wavefunctions for the protons
used in the zero temperature VMC calculation. Indeed, in CEIMC,
although electronic energies are calculated variationally, protons
are represented by a path-integral; the CEIMC results are
therefore more reliable.

\begin{table}
\begin{tabular}{|c|c|l|l|l|l|l|} \hline
 $r_s$   & $E_{tot}$ & $E_{kin}$ & $E_{pot}$ & $P$ & ZPE & $\gamma_L$ \\\hline
 0.8    &   -0.0573(7)      & 1.83(1)   & -1.89(2) & 80(1) & 0.036(1) & 0.18(1)    \\\hline
 1.0    &  -0.3477(6) & 1.208(2) & -1.556(2) & 20.02(6)  & 0.023(1)& 0.150(2) \\\hline
  1.2   &   -0.4637(4) & 0.873(2) & -1.337(2)  &5.51(8)  & 0.016(1) & 0.128(1) \\\hline
\end{tabular}
\caption{ Energies for $54$ dynamic protons and electrons at $T=0$ for various
densities
using VMC-TABC (averaged over $1000$ twist angles); total ($E_{tot}$), kinetic ($E_{kin}$) and potential energies ($E_{pot}$) are given in h/at, P is the pressure in Mbar, ZPE the zero point energies and $\gamma_L$ is the rms deviation divided by the nearest neighbor distance for a bcc lattice.
%Pressure=292.69*(2*E_kin-E_pot)/3.*density Mbar
}
\end{table}

%\section{Conclusions and discussion}

%\subsection*{Acknowledgments}
%Early aspects of the CEIMC algorithm were developed in
%collaboration with M. Dewing. We have the pleasure to thank
%J.P.Hansen and J.Kohanoff for useful discussions and for providing
%their CPMD data, and S.Scandolo for useful discussions. This work
%has been supported by a visiting grant from INFM-SezG and by
%MIUR-COFIN-2003. Computer time has been provided by NCSA
%(Illinois), PSC (Pennsylvania) and CINECA (Italy) through the INFM
%Parallel Computing initiative.

% \appendix
% costruct bibliography with BibTeX

\bibliography{template-num}
%\begin{thebibliography}{00}
% notes:
% \bibitem{label2}
% If there is a note, it should come last:
% \bibitem{label3} \note
% \end{subbibitems}
%\bibitem{}
%\end{thebibliography}

\end{document}